\documentclass[12pt]{article}

\usepackage{natbib}
\usepackage[hidelinks]{hyperref}
\usepackage{color}
\usepackage{multirow}

\usepackage{graphicx}
\usepackage{amsmath}
\usepackage{bbm}
\usepackage{comment}
\usepackage{mathtools}

\newcommand{\blind}{1}

\newcommand{\trans}{^{\top}}
\newcommand{\cov}{\textrm{cov}}
\newcommand{\bs}{{\boldsymbol{s}}}
\newcommand{\bh}{\boldsymbol{h}}
\newcommand{\mb}{\mathbf}

\newcommand{\bx}{\boldsymbol{x}}
\newcommand{\by}{\boldsymbol{y}}

\newcommand{\bH}{\mathbf{H}}
\newcommand{\bV}{\mathbf{V}}
\newcommand{\bB}{\mathbf{B}}
\newcommand{\bE}{\mathbf{E}}
\newcommand{\bY}{\mathbf{Y}}
\newcommand{\bW}{\mathbf{W}}
\newcommand{\bU}{\mathbf{U}}
\newcommand{\bSigma}{\mathbf{\Sigma}}

\newcommand{\bepsilon}{\boldsymbol{\epsilon}}

\begin{document}

\def\spacingset#1{\renewcommand{\baselinestretch}%
	{#1}\small\normalsize} \spacingset{1}

\if1\blind
{
	\thispagestyle{empty} \baselineskip=28pt \vskip 5mm
\begin{center} {\LARGE{\bf Forecasting High-Frequency Spatio-Temporal Wind Power with Dimensionally Reduced Echo State Networks}}
\end{center}

\begin{center}\large
Huang Huang\footnote[1]{
\baselineskip=11pt
Statistics Program,
King Abdullah University of Science and Technology, Thuwal, 23955-6900, Saudi Arabia. huang.huang@kaust.edu.sa, marc.genton@kaust.edu.sa
}, 
Stefano Castruccio\footnote[2]{
Department of Applied and Computational Mathematics and Statistics,
University of Notre Dame, Notre Dame, IN, 46556, USA. scastruc@nd.edu
\baselineskip=11pt } 
and Marc G. Genton\textsuperscript{1}
\end{center}

% \begin{center}\large
% Huang Huang\footnote[1]{
% \baselineskip=11pt
% Statistics Program,
% King Abdullah University of Science and Technology, Thuwal, 23955-6900, Saudi Arabia.}, Stefano Castruccio\footnote[2]{
% \baselineskip=11pt Department of Applied and Computational Mathematics and Statistics,
% University of Notre Dame, Notre Dame, IN, 46556, USA.} and Marc G. Genton\textsuperscript{1}
% \end{center} 

} \fi

\if0\blind
{
	\bigskip
	\bigskip
	\bigskip
	\thispagestyle{empty} \baselineskip=28pt \vskip 5mm

\begin{center} {\LARGE{\bf Forecasting High-Frequency Spatio-Temporal Wind Power with Dimensionally Reduced Echo State Networks}}
\end{center}

	\medskip
} \fi

\bigskip
\centerline{\today}
\bigskip
\bigskip

\begin{center}
Published in Journal of the Royal Statistical Society: Series C
\end{center}

\begin{abstract}
Fast and accurate hourly forecasts of wind speed and power are crucial in quantifying and planning the energy budget in the electric grid. Modeling wind at a high resolution brings forth considerable challenges given its turbulent and highly nonlinear dynamics. In developing countries, where wind farms over a large domain are currently under construction or consideration, this is even more challenging given the necessity of modeling wind over space as well. In this work, we propose a machine learning approach to model the nonlinear hourly wind dynamics in Saudi Arabia with a domain-specific choice of knots to reduce the spatial dimensionality.
Our results show that for locations highlighted as wind abundant by a previous work, our approach results in an $11\%$ improvement in the two-hour-ahead forecasted power against operational standards in the wind energy sector, yielding a saving of nearly one million US dollars over a year under current market prices in Saudi Arabia. 
\end{abstract}

	\noindent%
{\it Keywords:}  echo state network, Gaussian random field, machine learning, reservoir computing, space-time model
\vfill

\newpage
\spacingset{1.5} % DON'T change the spacing!

\addtolength{\textheight}{.5in}%

\section{Introduction}

The wide consensus of the increasingly negative effects of a warming climate caused by greenhouse gas emissions \citep{ipc14} has prompted the global community to seek alternative, carbon-free energy sources. While heterogeneous across countries, the increase in the usage of renewable resources worldwide has been steady over the past decades, and the aftermath of the current COVID-19 pandemic is foreseen to further accelerate this trend \citep{oecd20}. 

Saudi Arabia has the second-largest oil reserve in the world, and its internal energy portfolio is almost exclusively focused on fossil fuels. Worldwide, the country currently ranks sixth in oil consumption and has one of the highest per capita energy consumptions~\citep{bp20}. In an effort to align the country's targets with the Paris Agreement \citep{kin17} and reduce its carbon footprint, the Saudi Arabian government recently outlined ``Vision 2030'', a blueprint plan to promote an economically viable transition to renewable energies to reduce its dependence on oil. While solar energy has the greatest potential given the country's geographical position, the plan aims to also install 16 GW of wind capacity by 2030. Wind investment is strategic for the diversification and increased penetration of renewable energies given its high abundance, especially during night hours when solar power is not available. For a country with little to no installed capacity, the development of ``Vision 2030'' requires a preliminary understanding on where the turbines should be built and what factors are more influential in determining their sites. Very recent work provided some initial evidence. \citet{GTGCC2020} investigated various types of installation and maintenance costs and identified the optimal wind farm build-out under the present wind regimes; \cite{zha20} studied the resilience of the plan under a changing climate over the next thirty years; and \cite{che21} investigated the probability of disruption of wind operations due to extreme wind. 
One fundamental question must be addressed to provide a pathway for its practical implementation: What is the benefit of a reliable forecasting system, and what would be its economic advantage against standard operational forecasting systems? Such information would be instrumental to policymakers in their final decision regarding the optimal construction sites.

Modeling hourly wind fields (the resolution of interest in the energy market) for a country of one-fourth the size of the United States, with nonlinear dynamics implied by diverse geography comprising two coastal areas, mountain ranges, and sandy and rocky deserts, is a considerable challenge. Standard statistical approaches such as the Autoregressive Integrated Moving Average (ARIMA) models are effective in capturing a short-range (approximately) linear dependence; however, more flexible alternatives are necessary for nonlinear dynamics, and simple modifications such as fractional integration \citep{taq95} have been shown to be largely insufficient in capturing a more structured dependence in time. 

Machine learning approaches to time series, such as artificial neural networks, have been more successful in capturing nonlinear dynamics, owing to their flexibility and the possibility of defining a recursive structure (\textit{Recurrent Neural Networks}, RNNs) suitable for temporally resolved data. However, the inference for RNN is very challenging given the high dimensionality of the parameter space and the impossibility to directly apply the iterative gradient computations of traditional neural networks, which would lead to numerical instabilities \citep{D1992}. To achieve more stable and computationally affordable inference, \cite{Jaeger2001} proposed a new approach to RNN, i.e., the \textit{Echo State Networks} (ESN), which is predicated on the use of sparse, random matrices instead of dense unknown ones \citep{LJ2009}. \cite{MNM2002} independently developed a similar approach with the name \textit{Liquid State Machines} but with more sophisticated topological constraints motivated by neuroscience. This paradigm is now commonly referred to as \textit{Reservoir Computing} when inputs of the neural networks are mapped to high-dimensional hidden reservoir states through fixed nonlinear dynamics. These hidden reservoir states consider sequential linkages and thus allow for a nonlinear transformation of the input history. Despite the use of reservoir states being ideally suited as a baseline for statistical modeling, the machine learning community has mostly focused on this class of methods as a means to ease the computational burden. 

Recently, a series of seminal works \citep{MW2017,mcd19_a,mcd19_b} on the ESN framework was proposed in the statistical literature to forecast Pacific sea surface temperature and the United States soil moisture.
The Empirical Orthogonal Functions (EOF) method, or equivalently, the Principal Component Analysis (PCA), is used to reduce the spatial dimensionality. 
The leading orthogonal basis functions, which are the leading eigenvectors of the spatial covariance matrix of the space-time data, capture the main modes of spatial variability.
Therefore, they reduce the dimensionality and preserve the primary spatial variability structure when using the leading orthogonal basis functions with a much smaller number than the original number of locations to represent the entire spatial field.
In this work, however, we propose a new ESN framework that describes the dynamics of a spatial field through a collection of knots, sampled at both a grid and fixed points over areas of complex patterns such as rugged terrains (see Figure \ref{fig:knots}). The entire field is then reconstructed with a spatial interpolation approach, thereby allowing for a full spatio-temporal forecast without a computationally prohibitive ESN. Comparison with the EOF-based ESN, ARIMA, a spatio-temporal random process model, and the persistence forecasting (i.e., assuming the variable of interest does not change from the last observed value) shows that the prediction under our proposed ESN framework is more accurate.
Furthermore, we offer a comprehensive simulation study investigating how and to what extent the ESN can capture the data dynamics and yield predictive improvements against other methods when the underlying simulated process has a controlled level of nonlinearity. 

The manuscript proceeds as follows: Section \ref{sec:data} presents the wind data; Section \ref{sec:model} introduces the ESN and spatial model; Section \ref{sec:inf} discusses the inferential and forecasting approach; Section \ref{sec:forc} compares the forecasting from our model to several operational alternatives in the wind energy market; Section \ref{sec:power} presents the computation of the wind power forecast and discusses and quantifies the economic advantages of our method; and Section \ref{sec:conc} provides the conclusions of the study. 

Implementation of our proposed ESN for the wind speed forecasts in Saudi Arabia can be found at \url{https://github.com/hhuang90/KSA-wind-forecast}.

\section{Wind Data}\label{sec:data}

While developed countries can perform forecasting with publicly available, continuously updated high-resolution simulations routinely used for trading energy options (e.g., the High-Resolution Rapid Refresh~\citep{ben16} in North America), developing countries lack such data products. To investigate the wind speed and subsequent wind power potentials in Saudi Arabia, \cite{GTGCC2020} performed hourly high-resolution simulations from the Weather Research and Forecasting model \citep[WRF;][]{S2008} driven by the operational high-resolution European Centre for Medium-Range Weather Forecast model. WRF was simulated from 2013 to 2016, given the simultaneous availability of wind speed from meteorological masts for validation. The simulation was resolved on a regular planar grid with a fixed resolution from the Lambert Conformal Conic projection. 
Four simulations with different model setups were performed. For this study, we choose the one closest to the observations, which relied on the Mellor–Yamada–Janji{\'c} planetary boundary layer scheme~\citep{MY1982,J2001,GRD2015} and had a six-kilometer resolution, and we consider the hourly near-surface wind speed. For practical purposes, we exclude islands over the Red Sea (Tiran Island, Sanafir Island in the Red Sea Project Lagoon, and Farasan Island), for a total of $53,333$ locations.
\begin{figure}[tp!]
	\centering
    \includegraphics[width = 0.8\textwidth]{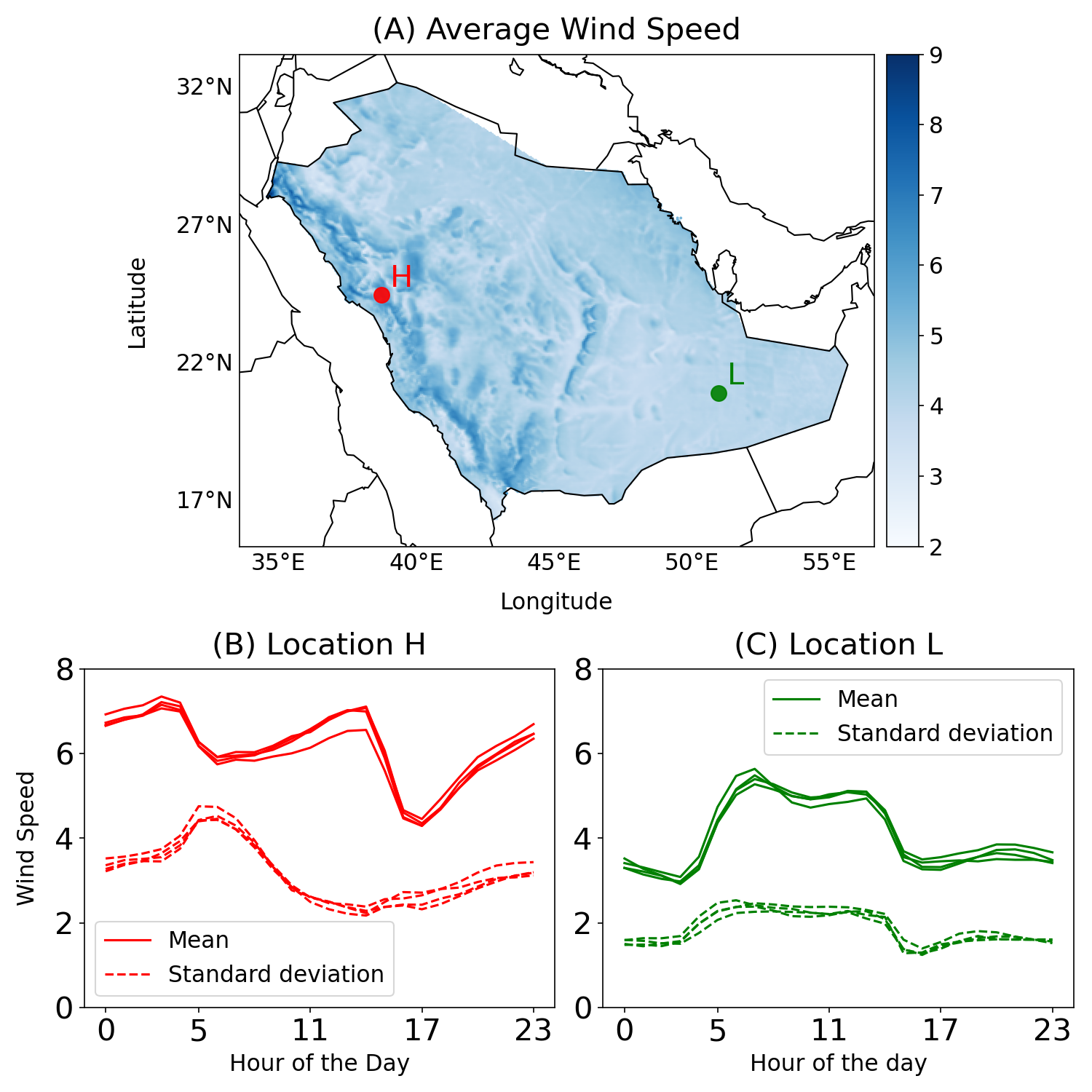}
	\caption{
		(A): Average wind speed in Saudi Arabia from 2013 to 2016, with two chosen locations illustrating high (H) or low (L) wind speed. (B) and (C): Mean and standard deviation of the hourly wind speed at each hour of the day in each year at locations H and L, respectively. Each of the four curves indicates one year from 2013 to 2016. Unit: m/s.
	    \label{fig:data}
	}
\end{figure}
Figure~\ref{fig:data} (A) depicts the average hourly near-surface wind speed from 2013 to 2016, showing generally higher values in the west due to mountain ranges. While abundant in wind, these regions are generally unsuitable for building wind farms due to the high installation costs implied by the rough terrain. 
The area in northwest Saudi Arabia near the Gulf of Aqaba is the construction site of NEOM City~\citep{Farag2019}, and data indicate its suitability for wind energy harvesting, as also discussed in~\cite{GTGCC2020}. We select two locations with a high or low average wind to display the diurnal temporal variability both in terms of mean and standard deviation in Figure~\ref{fig:data} (B) and (C).
We observe that the wind speed annual variability is low in the Arabian Peninsula, as also highlighted in \citet{CCGC2018}, hence lending support for the extension of the results in this work beyond the simulation years. WRF was validated by the observations at ten sites within the King Abdullah City for Atomic and Renewable Energy monitoring network through various assessment metrics in~\cite{GTGCC2020}.

\section{Models}\label{sec:model}

Throughout this work, we denote by $Z_t(\bs_i)$ the near-surface wind speed at location $\bs_i$ and time $t$, where $t$ takes values from $\{0,\ldots, T\}$ indicating the number of hours after 00:00 Jan.~1, $2013$ (Feb.~29 in the leap year of 2016 is removed for simplicity, but could be easily incorporated with more general harmonics in Equation~\eqref{eq:harmonic}), and $\bs_i,~i=1, \ldots, n=53,333$, is the data location in Saudi Arabia.

\subsection{Mean Structure}\label{subsec:online-problem}

We observe the periodic inter-daily and multi-daily patterns in the wind speed, which are explainable by both the daily land-ocean heat fluxes and mesoscale patterns of recurring winds in the Arabian Peninsula; see Figure~\ref{fig:data} for the daily patterns. For each location $\bs_i, i = 1,\ldots, n$, we assume that the dynamics of $\sqrt{Z_t(\bs_i)}$ depend on harmonics:
\begin{equation}
\label{eq:harmonic}
\sqrt{Z_t(\bs_i)}= \beta_0(\bs_i) + \sum\limits_{k=1}^{K} \left\{\beta_{k,1}(\bs_i)\cos\Big(\dfrac{2\pi t}{T_k}\Big) + \beta_{k,2}(\bs_i)\sin\Big(\dfrac{2\pi t}{T_k}\Big) \right\}+ \gamma(\bs_i)Y_t(\bs_i),
\end{equation}
where $K$ is the number of harmonics, $\beta_0(\bs_i)$ is the location-wise intercept, $\beta_{k,1}(\bs_i)$ and $\beta_{k,2}(\bs_i)$, $k = 1, \ldots, K$, are the location-wise coefficients for each harmonic with period $T_k$, and $\gamma(\bs_i)$ is the location-wise scaling parameter such that the residual $Y_t(\bs_i)$ has unit variance at each location. According to the discrete Fourier transform diagnostics illustrated in Figure~\ref{fig:DFT} in the Supplementary Material, there are $K=5$ significant harmonics at periods of one year, half a year, one day, twelve hours, and eight hours. We consider the square root because the wind speed is generally right-skewed due to occasional wind gusts. The choice of this particular transformation to normality is justified by well-established literature on wind modeling~\citep[e.g.,][]{G2002}, as well as our diagnostics in Figure~\ref{fig:qqplot} in the Supplementary Material. 

The parameters $\beta_0(\bs_i)$, $\beta_{k,1}(\bs_i)$, and  $\beta_{k,2}(\bs_i),~i = 1, \ldots, n,~k = 1,\ldots,K$, are estimated by linear regression independently across sites using the data from 2013 to 2015 (training data), a task that can be performed in parallel across locations. The scaling parameter $\gamma(\bs_i), ~i = 1, \ldots, n$, is also estimated site by site so that the resulting $Y_t(\bs_i)$ is a zero-mean spatio-temporal residual process with a structure specified in the following sections.

\subsection{Temporal Dynamics}
\label{subsec:temporal}
The hourly residual wind $Y_t(\bs_i),~i=1,\ldots,n$, is bound to show a highly nonlinear behavior that traditional time series approaches, such as ARIMA or more general linear space-time models, may fail to capture. The inability of traditional methods to capture the non-linearity of spatio-temporal dynamics~\citep{MW2017} necessarily translates into sub-optimal forecasts, as will be illustrated in Section~\ref{sec:forc}. We propose  an ESN model with nonlinear dynamics generated by a state space with neural network dynamics. We will introduce the model with the standard statistical terminology, but we also put in parenthesis the equivalent machine learning terms and use them henceforth. 

Quadratic interactions have been shown to be effective in characterizing the nonlinear dynamical evolution in many physical processes~\citep{WH2010,GW2014}. 
In this work, and consistently with~\cite{MW2017}, we propose to model the relationship between the data (\textit{output}) of the wind speed residual at time $t$ and all locations,
$\by_t:=\big(Y_t(\bs_1),\ldots,Y_t(\bs_n)\big)\trans$,
and the $n_h$-dimensional latent variables (\textit{hidden states} or \textit{reservoir states}), $\bh_t$,
by a quadratic ESN:
\begin{equation}
\label{eq:ESN-outputs}
\by_t = \bV_1\bh_t + \bV_2(\bh_t\odot\bh_t)+\bepsilon_t,
\end{equation}
where $\odot$ is the Hadamard (element-wise) product, $\bV_1$ and $\bV_2$ are weight matrices whose entries are unknown, and $\bepsilon_t$ is the error at time $t$, assumed to follow a zero-mean multivariate normal distribution. Here, $\bV_1$ and $\bV_2$ are estimated by least squares, as will be shown later in Section~\ref{subsubsec:est_out}, and are independent from the spatial covariance matrix of $\bepsilon_t$. The uncertainty of the estimated $\by_t$ relies on the covariance of $\bepsilon_t$, but we will show in Section~\ref{subsec:forecasting} that the prediction uncertainty can be calibrated with the ESN mean, by developing an approach similar to \cite{BC2021}.

The dynamics of the reservoir state vector $\bh_t$ is formulated as 
\begin{equation}
\label{eq:ESN-inputs}
\bh_t = \phi \mb{f}\left(\dfrac{\delta}{| \lambda_\bW |} \bW \bh_{t-1}+\bU \bx_t\right) + (1-\phi)\bh_{t-1}.
\end{equation}
The vector $\bx_t$ represents all the covariates (\textit{input}), and here
$\bx_t:=(1,\by_{t-1}\trans,\ldots,\by_{t-m}\trans)\trans$, i.e., we consider the intercept and the lagged wind speed residuals up to $m$ time steps before;
$\mb{f}$ is the element-wise activation function (here we choose the commonly used hyperbolic tangent function, but other choices are possible such as the rectified linear unit \citep{goo16}); $\lambda_\bW$ is the maximal absolute eigenvalue of $\bW$;
$\delta$ is a tuning parameter to scale $\bW$;
and $\phi$ is the leaking rate to control the speed of the dynamics. In summary, the temporal evolution of $\bh_t$ is controlled by two parts. The first one is a nonlinear evaluation of a linear combination of $\bh_{t-1}$ and $\bx_t$, and the second is a memory term that allows for long-range dependence. 

The $n_h\times n_h$-dimensional weight matrix $\bW=[w_{ij}]$ controls the interaction among reservoir states with time evolution, and the $n_h\times (mn+1)$-dimensional weight matrix $\bU=[u_{ij}]$ determines the mapping from the inputs to the reservoir states. The key difference between the ESN and the standard RNN lies in the structure of $\bW$ and $\bU$: an ESN assumes that these matrices are random with a controlled level of sparsity, while an RNN assumes full matrices with unknown fixed entries. Therefore, in the ESN, the interaction of each pair of reservoir states, as well as each pair of inputs and reservoir states, is assumed to exist with a given probability. The sparsity in both matrices prevents overfitting and dramatically reduces the model dimensionality. Specifically, the entries of $\bW$ and $\bU$ are generated as follows:
\begin{equation}
\label{eq:ESN-sparsity}
\begin{array}{c}
w_{ij} \mid \gamma^w_{ij} \sim \gamma^w_{ij} \text{Unif}(-a_w,a_w), \gamma^w_{ij}\sim \text{Bern}(\pi_w),\\
u_{ij} \mid  \gamma^u_{ij} \sim \gamma^u_{ij} \text{Unif}(-a_u,a_u), \gamma^u_{ij}\sim \text{Bern}(\pi_u),
\end{array}
\end{equation}
where $a_w$ and $a_u$ control the magnitude of nonzero entries in $\bW$ and $\bU$, respectively, and $\pi_w$ and $\pi_u$ determine the expected sparsity. 

An extensive literature in neural networks has proposed stochastic optimization methods, the most common being the random initialization of the weight matrices in the learning algorithm~\citep{YS2019}. ESNs are fundamentally different, as the weight matrices $\bW$ and $\bU$ are instead randomly generated from a given distribution; once drawn, they are fixed and do not need to be estimated. The randomness of $\bW$ and $\bU$ results in an ensemble of predictions, and its mean can be used as point prediction given its reduced variance with respect to a single ensemble member. 

\subsection{Spatial Modeling}
\label{subsec:spat}
The proposed ESN model could in principle be used to directly produce forecasts of $\by_t$; however, this would imply the use of input comprising a spatially referenced vector of $n=53,333$ points at each time $t$, and consequently a high-dimensional vector of reservoir states $\bh_t$, matrices $\bV_1, \bV_2$ in Equation~\eqref{eq:ESN-outputs} and matrices $\bW,\bU$ in Equation~\eqref{eq:ESN-inputs} to capture the dynamics. A brute-force approach disregarding the spatial information would therefore be computationally intractable. 

One commonly used approach to reduce the size of spatially referenced data in geosciences is using the EOF (or equivalently PCA) method. This approach allows to capture the main modes of spatial variability; for example, \cite{GW2014} used the first ten EOFs to represent the sea surface temperature on a region in the Pacific Ocean comprising of more than two thousand locations. In this work, we propose a stochastic approach with a Gaussian random field to model the spatial data. After fitting the Gaussian random field with all the available data, we consider a set of knots and use them to represent the entire process. Results in Section~\ref{subsec:forecast_wind} show that this modeling choice results in better predictive performances than EOFs.

We assume a zero-mean Gaussian random field to model $Y_t(\bs)$ at each time $t$, i.e., $Y_t(\bs)\sim \textrm{GP}\big(0,C(\cdot,\cdot)\big)$.
To reflect the spatially varying dependence structure dictated, among others, by the different topographical features in Saudi Arabia, we use a nonstationary Mat\'ern covariance function~\citep{Paciorek2006} for $Y_t(\bs)$. More specifically, for any pair of locations $(\bs,\bs')$, the covariance is 
\begin{equation}
\label{eq:matern}
C(\bs,\bs') = \sigma(\bs)\sigma(\bs')
R^{NS}\big(\bs, \bs^\prime; \bSigma(\cdot),\nu(\cdot)\big)
+\tau^2(\bs)\mathbbm{1}_{\{\bs=\bs'\}},\\
% &=& \sigma_t^2\dfrac{2^{1-\nu_t}}{\Gamma(\nu_t)}\left(\sqrt{2\nu_t}\dfrac{\|\bs-\bs'\|}{\rho_t}\right)^{\nu_t} K_{\nu_t}\left(\sqrt{2\nu_t}\dfrac{\|\bs-\bs'\|}{\rho_t}\right) + \tau_t^2\mathbbm{1}_{\{\bs=\bs'\}}.\\
\end{equation}
% \begin{sloppypar}
where $\sigma^2(\bs)$, $\nu(\bs)$, and $\tau^2(\bs)$ are the spatially varying partial sill, smoothness, and nugget parameters, respectively, $\bSigma(\bs)$ is the spatially varying covariance matrix of a Gaussian kernel centered at $\bs$, 
\[
\begin{array}{rl}
R^{NS}\big(\bs, \bs^\prime; \bSigma(\cdot),\nu(\cdot)\big) = &
\lvert\bSigma(\bs) \rvert^{\frac{1}{4}} 
\lvert\bSigma({\bs^\prime}) \rvert^{\frac{1}{4}} 
\left\vert \{\bSigma(\bs) + \bSigma(\bs^\prime)\}/{2} \right\vert^{-\frac{1}{2}} \times \\
&
R^{S}\big( 
\bs - \bs^\prime;
\{\bSigma(\bs) + \bSigma(\bs^\prime)\}/{2},
\{\nu(\bs) + \nu(\bs^\prime)\}/{2}
\big)
\end{array}
\]
is the nonstationary Mat\'ern correlation function, 
$
R^{S}\big( \boldsymbol{h}; \bSigma, \nu \big)
= 
2^{1-\nu}
\Gamma(\nu)^{-1}
(\boldsymbol{h}^\top\bSigma^{-1}\boldsymbol{h})^{\nu/2}\allowbreak
K_{\nu}\big((\boldsymbol{h}^\top\bSigma^{-1}\boldsymbol{h})^{1/2} \big)
$ is the stationary Mat\'ern correlation function~\citep{Stein1999}, $K_\nu(\cdot)$ is the modified Bessel function of the second kind with order $\nu$, and $\mathbbm{1}_{\{\}}$ is the indicator function. 
We use the \textsf{R}~\citep{R2019} package \textsf{convoSPAT}~\citep{MC2017} to fit the nonstationary random field with $Y_t(\bs_i)~i=1,\ldots,n$, in 2015.

\begin{figure}[ht!]
	\centering
	\includegraphics[width=\linewidth]{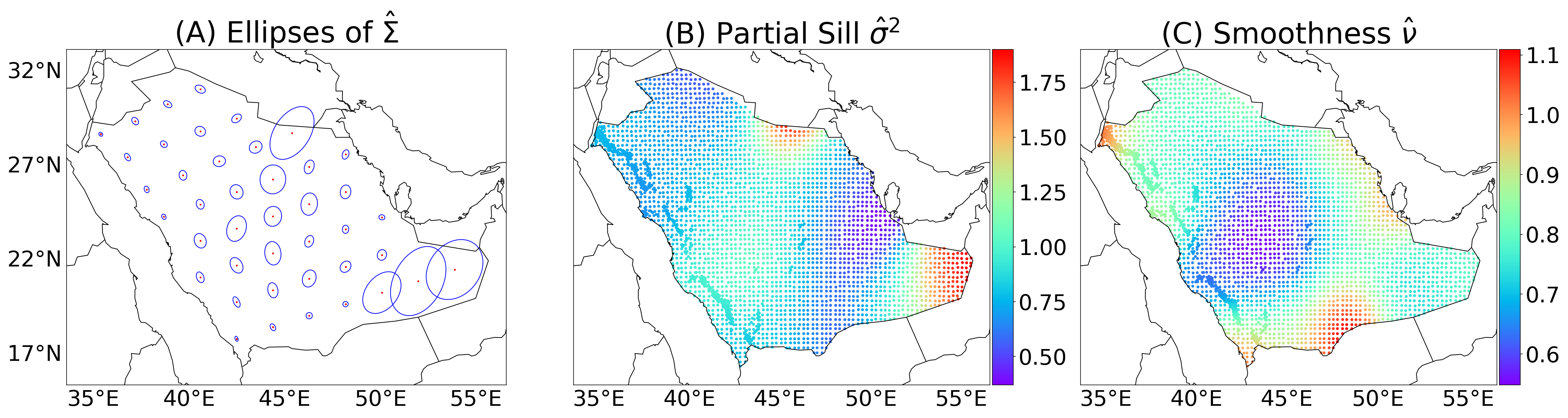}
	\caption{(A): Ellipses of estimate $\hat\bSigma(\cdot)$ at 42 mixture components (locations indicated as the red dot). (B) and (C): Spatially varying partial sill estimate $\hat\sigma^2(\cdot)$ and smoothness estimate $\hat\nu(\cdot)$ in the mixture component model.}
	\label{fig:ns_model}
\end{figure}

The package \textsf{convoSPAT} uses a mixture component model for $\sigma^2(\cdot)$, $\nu(\cdot)$, $\bSigma(\cdot)$ and $\tau^2(\cdot)$ and locally estimates the mixture parameters. We place 42 mixture components spanning the entire spatial domain as illustrated in Figure~\ref{fig:ns_model} (A). Our results suggest that the nugget estimate $\hat\tau^2(\cdot)$ is negligible, and the spatially varying partial sill estimate $\hat\sigma^2(\cdot)$ and smoothness estimate $\hat\nu(\cdot)$ are shown in Figure~\ref{fig:ns_model} (B) and (C).
The covariance of the Gaussian kernel characterizes anisotropy and the dependence scale, and we show in Figure~\ref{fig:ns_model} (A) the isocurves of the estimate $\hat\bSigma(\cdot)$ at the mixture components.

We adopt a moderately large set of knots across the spatial domain, and then model their space-time dynamics with the ESN. Once forecasts on the knots have been provided, spatial interpolation (\textit{kriging}) is used to produce spatial predictions. A formal comparison between the approximate process using knots and the true process has been studied, for example, in \citet{Titsias2009} and \citet{MHTG2016} using the Kullback-Leibler divergence. 

\begin{figure}[ht!]
	\centering
	\includegraphics[width = 0.7\textwidth]{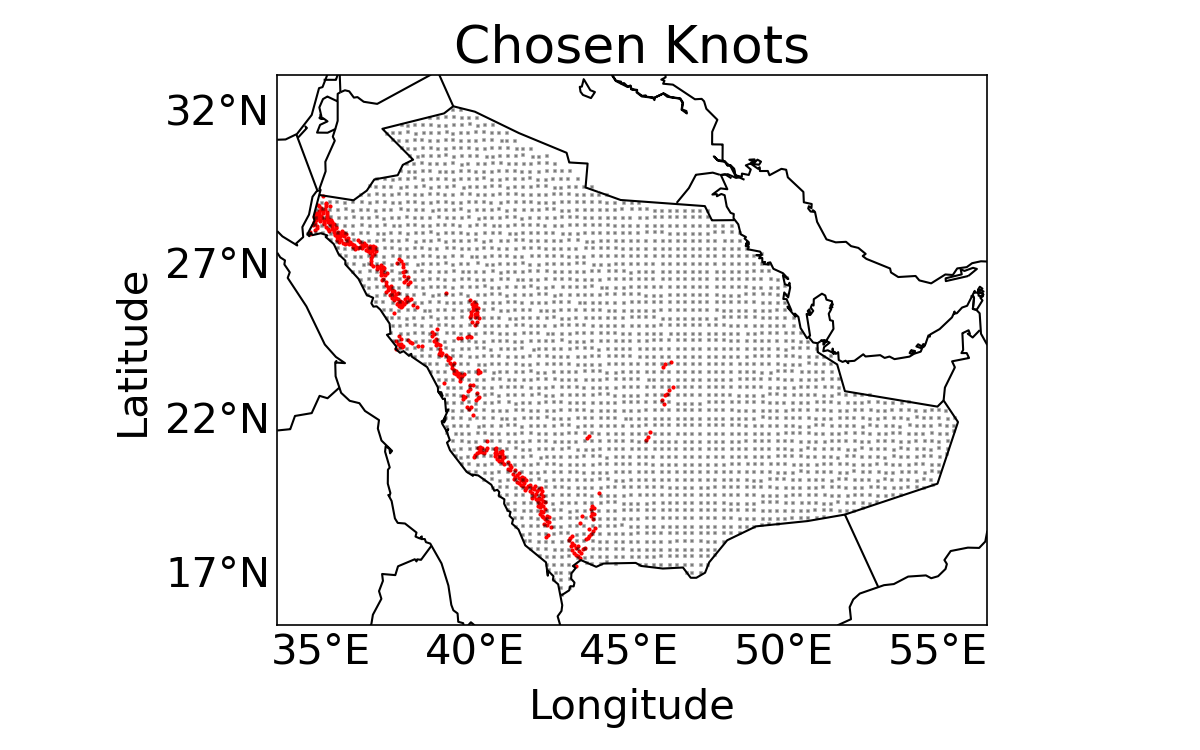}
	\caption{Overview of the $n^\ast=3,173$ chosen knots: 2,756 knots closest to a $0.25^\circ$-resolution regular grid are shown in light gray,
	and 417 knots with an average wind speed from 2013 to 2015 that is greater than 6 m/s are shown in red.}
	\label{fig:knots}
\end{figure}

Our chosen knots consist of two sets. The first set comprises 2,756 data locations closest to a $0.25^\circ$-resolution regular grid across the entire domain, while the second set is chosen at locations with a high wind power potential. We calculate the average wind speed from 2013 to 2015 and focus on the locations with values greater than 6 m/s, which is approximately the $98\%$ quantile. 
A set of 417 locations are then selected, such that no pairs have differences less than $0.005^\circ$ in longitude or latitude.
Figure~\ref{fig:knots} shows a total number of $n^\ast = 3,173$ chosen knots, which we denote by $\{\bs^\ast_1, \ldots, \bs^\ast_{n^\ast}\}$.
The output $\by_t$ is then replaced by
$\by_t^\ast:=\big(Y_t(\bs^\ast_1),\ldots,Y_t(\bs^\ast_{n^\ast})\big)\trans$.
The input $\bx_t$ is substituted accordingly with
$
\bx_t^\ast:=\left(1,{{\by_{t-1}^\ast}\trans},\ldots,{{\by_{t-m}^\ast}\trans}\right)\trans,
$
and the reservoir states $\bh_t$ in Equation~\eqref{eq:ESN-inputs} are updated by plugging in $\bx_t^\ast$ as
$
\bh_t = \phi \mb{f} \left(\frac{\delta}{|\lambda_\bW|} \bW \bh_{t-1} +\bU \bx_t^\ast\right) + (1-\phi)\bh_{t-1}.
$
\section{Inference and Forecasting}\label{sec:inf}

This section discusses both the inference and the forecasting approach. In Section~\ref{subsec:forecasting} we show how forecasting is performed given all parameter estimates and then in Section~\ref{sec:crossv} we elaborate on the inference procedure, where the tuning parameters are estimated via cross-validation based on the forecasting performance.

\subsection{Forecasting}
\label{subsec:forecasting}

Let the training period be $\{0,\ldots, T\}$ and the forecasting period be $\{T+1,\ldots,T_{\textrm{max}}\}$. In this work, we are not only interested in forecasts for one hour ahead,  but also for two and three hours ahead to conform our predictions with the majority of wind energy markets~\citep{HG2010}.
When we forecast $\by^\ast_{t+2}$ at time $t$, the observation $\by^\ast_{t+1}$ is required but unavailable, so we substitute the one-time-step forecast $\hat \by^\ast_{t+1}$ for $\by^\ast_{t+1}$ to forecast $\by^\ast_{t+2}$.
Section~\ref{sec:more-forecast} in the Supplementary Material formalizes the forecasts for multiple time-steps ahead.

At time $t$, once the forecast $\hat \by_{t+h}^\ast=\big(\hat Y_{t+h}(\bs^\ast_1),\ldots,\hat Y_{t+h}(\bs^\ast_{n^\ast})\big)\trans, h = 1,2,3$, at the knots is obtained, the wind speed residuals at all locations
$\hat \by_{t+h} = \big(\hat Y_{t+h}(\bs_1),\ldots,\hat Y_{t+h}(\bs_n)\big)\trans$ can be obtained via spatial interpolation, i.e., $\hat \by_{t+h} = \mathbf{K}^f\mathbf{K}^{-1} \hat \by_{t+h}^\ast$,
where $\mathbf{K}^f  = [k^f_{ij}]$ is an $n\times n^\ast$-dimensional covariance matrix with entries 
$k^f_{ij} = \cov\big(Y_{t+h}(\bs_i),Y_{t+h}(\bs^\ast_j)\big)$ and $\mathbf{K}  = [k_{ij}]$ is an $n^\ast\times n^\ast$-dimensional covariance matrix with entries 
$k_{ij} = \cov\big(Y_{t+h}(\bs^\ast_i),Y_{t+h}(\bs^\ast_j)\big)$.

As pointed out by \citet{EH2016}, averaging a set of unbiased estimators is an effective way to reduce estimator variance. \cite{Rougier2016} also investigated why the ensemble mean often yields a smaller Mean Squared Error (MSE) than the majority  of the ensemble members. Throughout this work, we use 100 ensemble members to forecast, each consisting of an independent random realization of $\bW$ and $\bU$. The mean of the 100 ensemble forecasts is then used as the final forecast. In order to quantify the uncertainty, an empirical calibration similar to that proposed in \citet{BC2021} is used here. An ESN is trained with the wind speed residuals $Y_t(\bs)$ from 2013 to 2014, and the forecast for 2015 (calibration period) is produced. At each location $\bs$, the quantiles of $Y_t(\bs) - \hat Y_t(\bs)$ in 2015 are estimated. Then, these quantiles will be used to build the forecast distribution for $\hat Y_t(\bs)$ in 2016.

\subsection{Inference}\label{sec:crossv}

\subsubsection{Estimating the output matrices}
\label{subsubsec:est_out}
Estimating $\bV_1$ and $\bV_2$ can be performed via least square regression. The large size of the output vector would potentially result in a high dimension of the reservoir state space. Therefore, it is necessary to penalize the coefficient estimation to reduce its variance and prevent overfitting.
If we denote 
\[
\bY = \left[\begin{array}{c}
{\by^\ast_1}\trans\\
\vdots\\
{\by^\ast_T}\trans\\
\end{array}\right],
\bH = \left[\begin{array}{c}
\bh\trans_1,(\bh_1\odot\bh_1)\trans\\
\vdots\\
\bh\trans_T,(\bh_T\odot\bh_T)\trans\\
\end{array}\right],
\bB = \left[\begin{array}{c}
\bV_1\trans\\
\bV_2\trans\\
\end{array}\right],
\bE = \left[\begin{array}{c}
\bepsilon\trans_1\\
\vdots\\
\bepsilon\trans_T\\
\end{array}\right],
\]
then, we can express Equation~\eqref{eq:ESN-outputs} in a matrix linear regression form as $\bY = \bH\bB+\bE$. We use ridge regression or Tikhonov regularization to estimate $\bB$ such that 
\[
\hat \bB = \arg\min_\bB(\|\bY-\bH\bB\|^2_\textrm{F} + \lambda \|\bB\|^2_\textrm{F}),
\] 
where $\|\cdot\|_\textrm{F}$ is the Frobenius norm and $\lambda$ is the ridge penalty.
The ridge estimator $\hat \bB$ has the following closed-form:
\begin{equation}
\label{eq:ridge}
\hat \bB=(\bH\trans \bH+\lambda \mathbf{I})^{-1}\bH\trans \bY.
\end{equation}
Once $\hat\bB$ is obtained, the forecast for one time-step ahead (and iteratively for any future time step) on the knots can be computed as follows:
\[
\hat\by_t^\ast = 
\hat\bB\trans
\left[ 
\begin{array}{c}
\bh_t\\
{\bh_t\odot\bh_t}
\end{array}
\right],
\]
for $t\in \{T+1,\ldots,T_\textrm{max}\}$. 

\subsubsection{Cross-validation}

Except for $\bV_1$ and $\bV_2$, several other parameters need to be estimated via cross-validation: the dimension of the reservoir states $n_h$, the number of lags $m$ in $\bx_t$, the leaking rate $\phi$ and the scaling parameter $\delta$ in Equation~\eqref{eq:ESN-inputs}, the magnitude and sparsity parameters $a_w,\pi_w,a_u$, and $\pi_u$ in Equation~\eqref{eq:ESN-sparsity}, and the ridge penalty $\lambda$ in Equation~\eqref{eq:ridge}. We consider $\by^\ast_{t}$ from 2013 to 2014, for a total of 17,520 hourly observations, as the training data to obtain the ridge estimator $\hat\bB$ in Equation~\eqref{eq:ridge}. Subsequently, $\by^\ast_{t}$ in 2015, for a total of 8,760 hourly observations, is used as the validation data. More specifically, at any time $t$ in 2015, $\by^\ast_0,\ldots,\by^\ast_t$ are available; we forecast $\by^\ast_{t+1}$, we compare it with the true validation data, and we then calculate the MSE at all knots and time points for the mean of 100 ensemble forecasts (see Section~\ref{subsec:discard} in the Supplementary Material for more details). We search over a large grid (shown in Table~\ref{tab:search_grid}) to obtain the optimal values of the parameters, see Table~\ref{tab:par}.

\begin{table}[ht!]
	\centering
	\caption{\label{tab:par} Estimated parameters in the ESN model via cross-validation. $\by^\ast_{t}$ from 2013 to 2014 is used as the training data set, and $\by^\ast_{t}$ in 2015 the validation data set.}
	\vspace{3mm}
	\begin{tabular}{|c|c|c|}
		\hline
		Parameter explanation & Notation & Estimate	 \\ \hline
		Number of reservoir states	& $n_h$					& 2,500 \\ \hline
		Input lag  &    $m$	                                & 1 \\ \hline
		Leaking rate  & $\phi$	                            & 1	\\ \hline
		Scaling matrix parameter for $\bW$ & $\delta$	        & 0.9	\\ \hline
		Ridge penalty & $\lambda$	                        & 0.15	\\ \hline
		Magnitude of entries in $\bW$ & $a_w$	                & 0.05	\\ \hline
		Magnitude of entries in $\bU$ & $a_u$	             & 0.01	\\ \hline
		Sparsity of $\bW$ & $\pi_w$	            & 0.1	\\ \hline
		Sparsity of $\bU$ & $\pi_u$	                        & 0.01	\\ \hline
	\end{tabular}
	
\end{table}

The large number of reservoir states indicates that a complex model is necessary to capture the nonlinearity of the multivariate time series with $n^\ast=3,173$ dimensions. The estimated ridge penalty avoids overfitting using the large reservoir space. The obtained optimal leaking rate is one (i.e., the reservoir states are fully updated by new activations at each time). Since we removed the low-frequency periodicities to the original wind speed (see Equation~\eqref{eq:harmonic}), the optimal leaking rate of one suggests a short-term dynamic in the residual field $Y_t(\bs)$. In our ESN, we use the lagged wind speed residuals as the input, and we observe that the optimal model only requires one lagged observation as the input at each time. The parameter $\delta$ determines the spectral radius of the matrix $\bW$ after scaling and influences the fading speed of the input in the hidden neurons. 
In practice, $\delta<1$ always guarantees the echo state property (i.e., the reservoir states are independent on the initial conditions after a sufficiently long time). %However, it is also observed that the echo state property often holds when $\delta\geq1$ because of the activation-squashing nonlinearities of the hyperbolic tangent functions~\citep{L2012}. Our obtained optimal value of $\delta$ indicates that even though only one lagged observation is directly needed at each time, the hidden neurons preserve long memory of the inputs (lagged wind speed residuals) by employing a moderately large spectral radius of $\bW$, 2.2. 
Our estimated value of $\delta$ indicates that even though only one lagged observation is necessary at each time, the hidden neurons preserve a long memory of the input (lagged wind speed residuals) by employing a moderately large spectral radius of $\bW$, 0.9.
The values of $a_w,\pi_w,a_u$, and $\pi_u$ are all comparatively small, implying a sparse collection of weak connections among reservoir states and between the reservoir states and the input.

\section{Forecasting  Evaluation}\label{sec:forc}

\subsection{Forecasting Results for Wind Speed}
\label{subsec:forecast_wind}

Once the parameters have been estimated in the training step from 2013 to 2015, we proceed to forecast $\by_t^\ast$ in 2016 and compare it with other common strategies for short-term wind forecasts. The calibration algorithm explained in Section~\ref{subsec:forecasting} is used to obtain the probabilistic forecast distribution in 2016. 
Table~\ref{tab:UQ-knots} shows the proportion of the true wind speed residuals $\by_t^\ast$ in 2016 falling into the associated prediction intervals, and we observe that overall they match the nominal levels closely.

\begin{table}[ht!]
	\centering
	\caption{\label{tab:UQ-knots} The mean proportion of the wind speed residuals $\by_t^\ast$ in 2016 falling into the associated prediction intervals (standard deviation across the $n^\ast$ knots is shown in parentheses).}
	\vspace{3mm}
	\begin{tabular}{|c|c|c|c|}
		\hline
		Prediction &  \multicolumn{3}{c|}{Prediction Interval Coverage} \\
		\cline{2-4}
		Interval & One hour ahead & Two hours ahead & Three hours ahead \\
		\hline
		$95\%$  &  $94.8\%(0.5\%)$  &  $94.6\% (0.6\%)$  &  $94.6\% (0.7\%)$ \\
        $80\%$  &  $80.7\%(1.0\%)$  &  $80.6\% (1.1\%)$  &  $80.6\% (1.2\%)$ \\
        $60\%$  &  $60.7\%(1.3\%)$  &  $60.3\% (1.3\%)$  &  $60.1\% (1.5\%)$ \\
		\hline
	\end{tabular}
	
\end{table}

We compare the ESNs with persistence forecasting and the ARIMA model. The persistence forecasting method uses the last observed values as estimates for the near future (i.e., $\hat\by_{t+3} = \hat\by_{t+2} = \hat\by_{t+1} = \by_t$). This method is very simple, but frequently used in practice because of its good performance for very short-term forecasting. Additionally, we propose an ARIMA approach, which assumes that at each knot $\bs^\ast_i, i=1,\ldots, n^\ast$, the time series of wind residual $Y_t(\bs^\ast_i)$ has the following model:
\[
\left(1-\sum^{p(\bs^\ast_i)}_{k=1}\psi_k(\bs^\ast_i)\mathcal{L}^k\right)
(1-\mathcal{L})^{d(\bs^\ast_i)}
Y_t(\bs^\ast_i) 
= 
c(\bs^\ast_i) +
\left(1+\sum^{q(\bs^\ast_i)}_{k=1}\theta_k(\bs^\ast_i)\mathcal{L}^k\right)
\varepsilon_t(\bs^\ast_i),
\]
where $p(\bs^\ast_i)$, $d(\bs^\ast_i)$, and $q(\bs^\ast_i)$ are the order of the autoregressive, differencing, and moving-average terms, respectively.
Here, $\mathcal{L}$ is the lag operator (e.g., $\mathcal{L}Y_t(\bs^\ast_i) = Y_{t-1}(\bs^\ast_i)$),
$\varepsilon_t(\bs^\ast_i)$ is the error term, 
and $c(\bs^\ast_i)$, $\psi_1(\bs^\ast_i),\ldots,\psi_{p(\bs^\ast_i)}(\bs^\ast_i)$ and $\theta_1(\bs^\ast_i),\ldots,\theta_{q(\bs^\ast_i)}(\bs^\ast_i)$ are unknown parameters at $\bs^\ast_i$. The optimal orders for the ARIMA model at each knot are independently optimized through cross-validation using the data from 2013 to 2015. 
For each time in 2016, the ARIMA model with the optimal parameters uses all the data before that time and computes the forecasts at each knot up to three hours ahead.

Table~\ref{tab:MSE-knots} summarizes the MSE results for $\by_t^\ast$ at all the $n^\ast$ knots and time points in 2016. The ESN forecasts outperform these from persistence and ARIMA, with better results as the lead time increases from one to three hours ahead.

\begin{table}[ht!]
	\centering
	\caption{\label{tab:MSE-knots} MSE for the forecasted $\by_t^\ast$ at all the $n^\ast$ knots and time points in 2016 by the ESN, ARIMA, and persistence methods.}
	\vspace{3mm}
	\begin{tabular}{|c|c|c|c|}
		\hline
		Forecast &  ~~~~ESN~~~~  & ~~ARIMA~~ & Persistence \\ \hline
		One hour ahead 	  &  {0.235}	&  	0.292			& 0.326\\
		Two hours ahead   &  {0.394}	&  	0.533			& 0.657\\
		Three hours ahead &  {0.508}	&  	0.689			& 0.920\\
		\hline
	\end{tabular}
	
\end{table}

\begin{figure}[ht!]
	\centering
	\includegraphics[width = \textwidth]{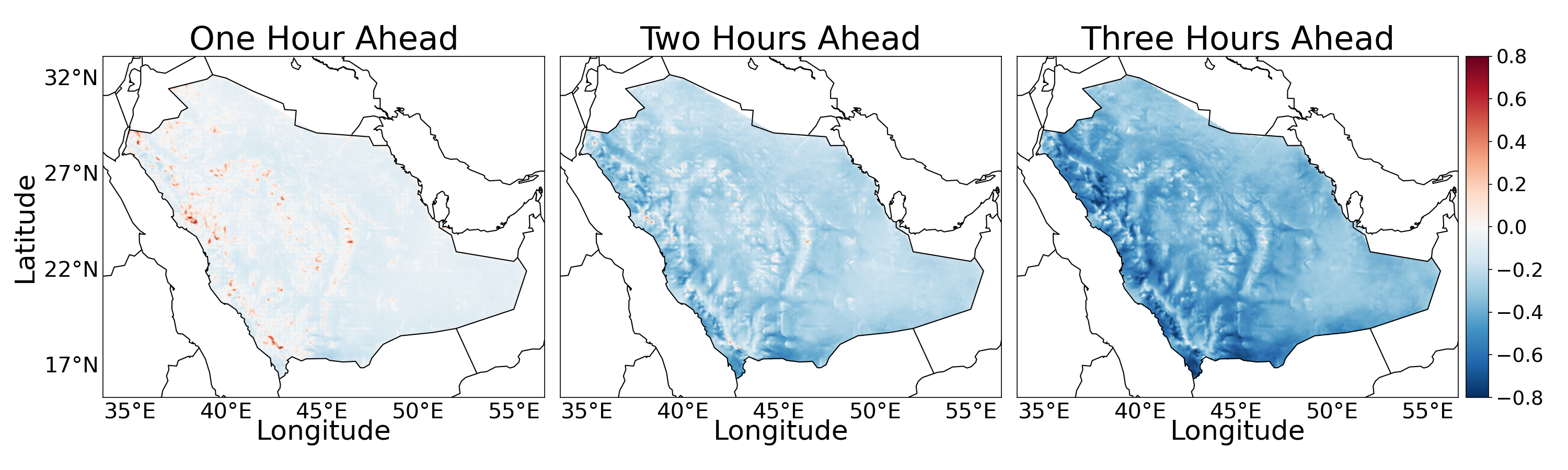}
	\caption{MSE for the forecasted $\by_t$ at each of the locations in 2016 by the S-ESN minus MSE for the persistence method.}
	\label{fig:MSE_difference}
\end{figure}

After forecasting $Y_t(\bs)$ at the $n^\ast$ knots, we produce a map at the $n$ locations by spatial interpolation from the inferred spatial Gaussian random field described in Section~\ref{subsec:spat}, and we denote this approach by S-ESN. We compare this method with ARIMA or persistence approaches trained on all the $n$ locations, a choice which puts the S-ESN at a competitive disadvantage, as it can learn the dynamics only on the knots. Table~\ref{tab:MSE} shows the outperformance of ESN over the ARIMA and persistence in terms of the average MSE, and in Figure~\ref{fig:MSE_difference} we show the maps of the MSE difference between S-ESN and persistence for one to three hours ahead. We observe smaller errors for our forecasts than the persistence forecasts at the majority of locations. We also provide probabilistic forecasts of $Y_t(\bs)$ at all locations in 2016 based on the calibrated quantiles from 2015. Table~\ref{tab:UQ-all} in the Supplementary Material shows the prediction interval coverage, and we observe that the uncertainty is well captured.

\begin{table}[ht!]
	\centering
	\caption{\label{tab:MSE} MSE for the forecasted $\by_t$ at all locations and time points in 2016 by the S-ESN, EOF-ESN, ARIMA, FRK, and persistence methods.}
	\vspace{3mm}
	\begin{tabular}{|c|c|c|c|c|c|}
		\hline
		Forecast & ~~S-ESN~~  & EOF-ESN  & ~ARIMA~ & ~~~FRK~~~ & Persistence						 \\ \hline
		One hour ahead 	  &  {0.276} & 0.364 & 0.302 & 0.316 & 0.335\\
		Two hours ahead   &  {0.424} & 0.570 & 0.548 & 0.554 & 0.670\\
		Three hours ahead &  {0.537} & 0.705 & 0.707 & 0.788 & 0.936\\
		\hline
	\end{tabular}
	
\end{table}

As mentioned in Section~\ref{subsec:spat}, the EOF approach is also commonly used to reduce dimensions~\citep{MW2017}. We denote this method by EOF-ESN, and we compare it with our method. We use $\by_t$ at all locations from 2013 to 2015 to calculate its covariance matrix and extract the first $n_{\text{EOF}}=3,173$ EOFs (principal components).
We then reduce the dimension of $\by_t$ to $n_{\text{EOF}}$, apply the ESN to the projected vector, compute forecasts up to three hours ahead, and recover the forecasted map at all locations by projecting the EOFs back into the original space. Results in Table~\ref{tab:MSE} indicate better forecast skills from the S-ESN, which relies on spatial information to reduce dimensionality, as opposed to EOF-ESN, a data-driven, non spatially informed method for dimension-reduction. We reiterate that in Table~\ref{tab:MSE}, the persistence forecasting method relies on past values at all locations, whereas S-ESN and EOF-ESN only model the data in a reduced space. Despite the use of a more limited set of data, even the EOF-ESN is competitive against persistence: for two and three hours ahead, it produces better forecasts.

Traditional spatio-temporal random process models can also be used to provide wind forecasts. However, given the high dimension in space and time, approximation techniques need to be applied for inference and prediction. We use the Fixed Rank Kriging \citep[FRK;][]{CJ2008} approach as one example for comparison. FRK proposes a spatio-temporal random effect model, where the spatio-temporal random process is decomposed by a series of basis functions, thus achieving dimensionality reduction and naturally accounting for nonstationarity. We use the recently developed \textsf{R} package \textsf{FRK}~\citep{ZC2021} to model $\by_t$, and the results are shown in Table~\ref{tab:MSE}. FRK yields more accurate forecasts than persistence and EOF-ESN for one and two hours ahead. We notice that the outperformance of FRK over EOF-ESN reduces as the lead time increases (FRK becomes worse than EOF-ESN for three hours ahead), indicating that the temporal dynamic may not be well characterized by the spatio-temporal process. For all three forecast horizons, S-ESN and ARIMA have consistently more accurate results than FRK. One possible reason for the outperformance of ARIMA over FRK is that ARIMA computes the location-wise forecast using all the available data directly but FRK uses basis functions for the spatio-temporal data, which implies some level of approximation.

In conclusion, we stress the remarkable result of our proposed S-ESN approach: by only relying on data at the knots (about $6\%$ of all the locations) S-ESN is able to outperform alternative methods, even for very short-term forecasts.

Once the wind speed forecasts are produced, we proceed to evaluate the associated predictions for wind power. While it may be interesting to perform a comparison over the entire spatial domain, the practical interest lies in the locations suitable for wind farm construction. In Section~\ref{sec:power}, we investigate the generated wind power for locations highlighted as suitable for wind farms by \citet{GTGCC2020}. It is also noteworthy that MSE may not be a perfect metric to assess wind speed or associated wind power due to the possibly asymmetric economic cost between under- and over-forecasts. The evaluation can be performed differently by replacing the MSE with a case-specific commercial loss function.

\subsection{Simulation Study on a Modified Lorenz 96 Model}
To better understand the ability of ESN to capture nonlinear dynamics, we also perform a simulation study from a model with a controlled level of nonlinearity in the temporal dynamics. We use one of the most popular models in numerical weather prediction, the Lorenz 96 model~\citep{Lorenz1996}, and we modify it by adding a factor to control the nonlinearity in the dynamics.
We have observed that ESN leads to the most accurate forecasts compared to the ARIMA, vector autoregression, and persistence methods. When the nonlinearity of the dynamics is stronger, the outperformance of the ESN is more significant. Details of the simulation study for the modified Lorenz 96 model is given in Section~\ref{subsec:Lorenz} in the Supplementary Material.

\section{Wind Power Forecasts in Saudi Arabia}\label{sec:power}

Once our forecasting method has been validated against existing alternatives, we proceed to produce the forecasts of wind power. From an operational perspective, the electricity produced by wind power cannot be stored and must be consumed once it is injected into the power grid. An accurate forecast of the wind power helps to plan the correct amount of additional electricity to be dispatched from other sources. Penalties or fines are applied in utility markets in case of unfulfilled commitment to provide an agreed amount of power. The typical forecast horizon for scheduling electricity transmission, allocating resources, and dispatching generated power is two to four hours \citep{GLWGA2006,HG2010}. 
Herein, we focus on two-hour-ahead wind forecasts. We study the wind power forecasts based on the optimal wind farm build-out in Saudi Arabia as shown in \cite{GTGCC2020} and assess how much power would be saved with our forecasts if all the identified wind farms were installed and in operation.

The two-hour-ahead forecast of $Y_t(\bs)$ is transformed into near-surface wind speed $Z_t(\bs)$ according to Equation~\eqref{eq:harmonic}. Since the wind speed is different at different heights, we convert the wind speed near the surface (10 meters above the ground) to the wind turbine hub height. The wind power law is commonly used to perform this conversion:
\[
Z^{(h)}_t(\bs) = Z_t(\bs)(h/10)^{\alpha(\bs,t)},
\]
where $h$ is the hub height and $\alpha(\bs,t)$ is the location- and time-specific factor. Traditionally, the factor $\alpha(\bs,t)$ is fixed as a constant for all locations and times, and the value 1/7 is usually used over open land surfaces~\citep{PB2011} and specifically over Saudi Arabia~\citep{REASAB2007,TCCG2019}.
However, \citet{CABGC} showed that if the location- and time-specific factor $\alpha(\bs,t)$ is independently estimated at each location from different profile heights, the recovered wind speed at the hub height is on average 36\% more accurate, so we use their model to convert the wind speed vertically. It is noteworthy that this vertical extrapolation model by the wind power law would lead to some errors. In this study, we do not take into account the uncertainty of $\alpha(\bs,t)$ for the probabilistic forecasts of wind power.

There are two turbine models chosen for the 75 identified optimal wind farms~\citep{GTGCC2020}: Nordex N131/3300 and GE Energy 2.75-100, with hub heights 84 and 75 meters above the ground, respectively.
We compute the wind speed for each wind farm at the turbine hub heights and then use the \textit{power curve} provided by the wind turbine manufacturers to convert the wind speed to electric power (see Figure~\ref{fig:power_curve}). Each curve has three critical points that divide the power curve into four zones. When the wind speed is less than the cut-in speed, the turbine motor does not rotate. When the wind speed is greater than the cut-in but less than the rated speed, the turbine motor produces more power with faster wind. Once the wind speed reaches the rated speed but not greater than the cut-out speed, the turbine keeps producing the maximal power output. If the wind speed goes beyond the cut-out speed, the turbine motor stops rotating to avoid damage and produces no power. Therefore, better wind speed forecasts do not necessarily yield better wind power forecasts, as their relationship is not strictly monotonic.

\begin{figure}[ht!]
	\centering
	\includegraphics[width = .8\textwidth]{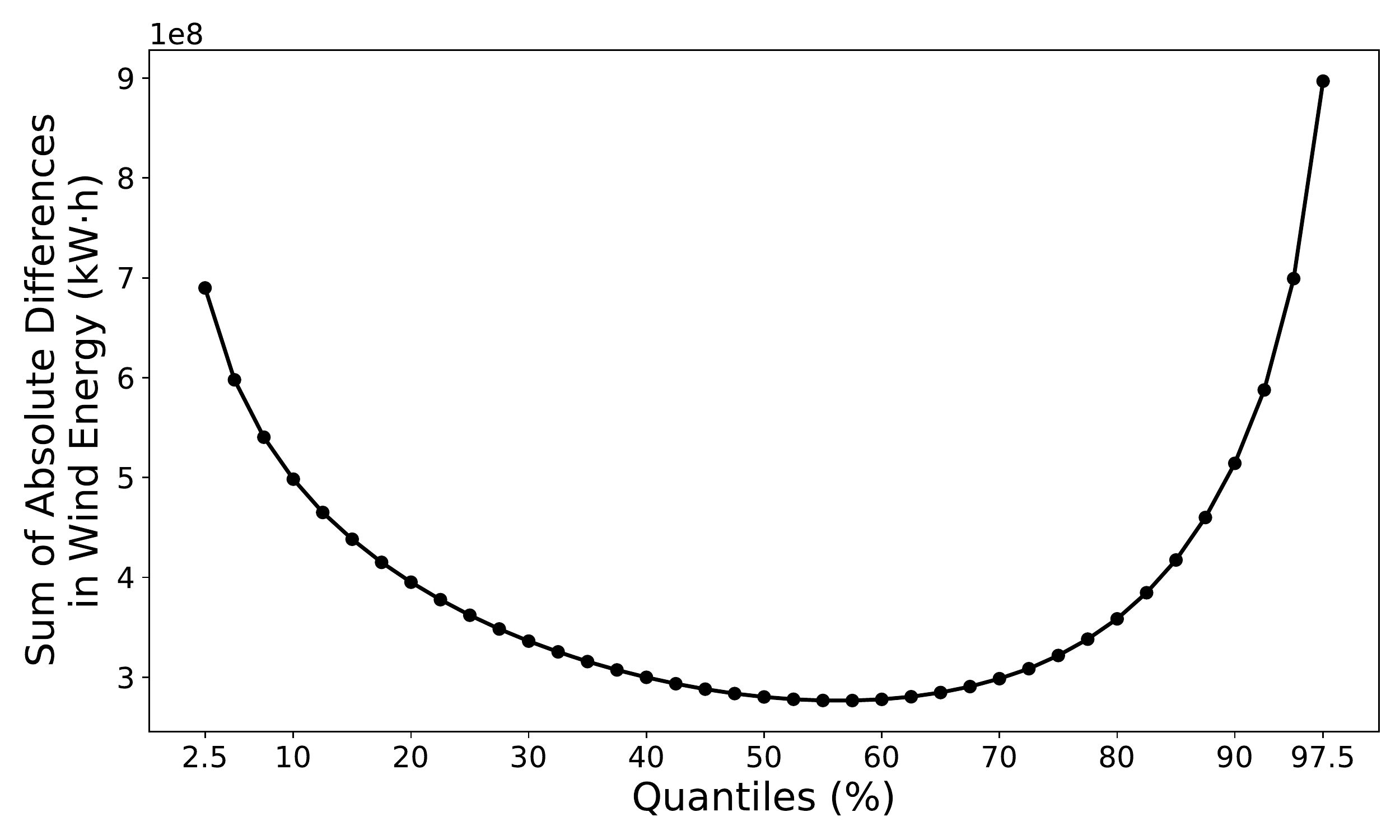}
	\caption{The annual sum of the absolute differences between the S-ESN forecast quantiles and the truth in wind energy during 2016 at all the 75 wind turbines combined.}
	\label{fig:power_quantile}
\end{figure}

We also convert the forecasted near-surface wind speed by the ARIMA and persistence methods as well as the true wind speed at the 75 wind farms, and we compare their error in wind power forecasts  against our approach. Note that the economic cost may be asymmetric for under- and over-estimation of wind power. Under-forecast occurrence means that the power grid has ordered more electricity than needed, which results in a power surplus. Then, the power generation must be reduced, and it is usually more expensive than the opposite case. This asymmetry is case-specific and can vary in different countries, and for simplicity, we do not distinguish the direction of the wind power forecast error in this work. We calculate the absolute difference between the S-ESN (similarly, the ARIMA or persistence) wind power forecasts and the true power for each hour in 2016. The annual sum of the absolute differences in wind energy for all the 75 wind turbines is $2.77\times 10^8$ kW$\cdot$h for S-ESN, $3.05\times 10^8$ kW$\cdot$h for ARIMA, and $3.12\times 10^8$ kW$\cdot$h for persistence. Thus, we obtain a $9\%$ improvement against the ARIMA and an $11\%$ improvement against the persistence forecasts. The bid price in the electricity market varies based on many factors (demand and supply, providers, etc.). Modeling the imbalance price is generally difficult and out of the scope of our work; see \cite{ZG2012} and \cite{Pinson2013} for some discussion on the operational management in an electricity market. To gain some insights, using a medium bid price of 0.025 US dollars per kW$\cdot$h nowadays in Saudi Arabia as a reference, the forecasts from our approach could yield a saving of nearly one million US dollars over a year against the persistence forecasts. When probabilistic forecasts are considered, Figure~\ref{fig:power_quantile} shows the whole-year difference between the S-ESN forecast quantiles and the truth in wind energy. The difference curve around center quantiles is quite flat. Specifically, the differences from $40\%$ to $70\%$ quantiles vary from $2.77\times 10^8$ kW$\cdot$h to $3\times 10^8$ kW$\cdot$h, and the maximum difference from $20\%$ to $80\%$ quantiles reaches $3.95\times 10^8$ kW$\cdot$h. It is also observed that the difference is inflated drastically for extreme quantiles. Finally, we notice that the wind energy forecast error is asymmetric in quantiles, owing to the characteristics of the power curve illustrated in Figure~\ref{fig:power_curve}.

\section{Discussion and Conclusion}\label{sec:conc}

In this work, we have introduced a novel ESN framework for spatio-temporal wind data. The approach is shown to have superior predictive performances than the operational standards in the wind energy market, contributes to the spatio-temporal model literature, and has practical implications for the wind energy sector in Saudi Arabia.

The proposed model uses the high dimensionality of neural networks to capture the nonlinear dynamics of hourly wind but provides two solutions to ease the computational burden of the model. 
The paradigm of reservoir computing allows achieving sparsity in the RNN matrices, thereby dramatically reducing the parameter space; the use of spatial (problem-informed) knots allows reducing the spatial dimensionality while preserving a larger amount of information than previous approaches. 
Although our method shows high predictive power, it represents only one of the many possibilities available to bridge the gap between machine learning methods and spatio-temporal statistics.
Many possible alternatives linking state-of-the-art geospatial models and neural network forecasting techniques are still unexplored.

This work presents a fast and efficient approach to hourly wind forecasting over large areas. A more detailed quantification of the economic gains from an improved forecast is reliant upon a market model for trading and bidding energy. The energy sector of Saudi Arabia is currently under full governmental control; hence, the integration of our forecasting approach would require considerable speculation on the market model that will be employed in the future. 
Therefore, we choose not to pursue this route but rather opt for a simpler comparison against operational standards, such as persistence, under the current market prices in Saudi Arabia. 
Some of our current studies focus on engaging with local leaders in the energy sectors to detail the range of possible future market models that could be integrated with our approach. 
In more immediate terms, this work will provide additional information on where to build turbines aside from the wind abundance and construction and operation costs in \cite{GTGCC2020}. 
Finally, while this work is reliant upon (validated) high-resolution WRF simulations, it represents a template to assess wind predictions over a large spatial area for developing countries in the Arab Gulf, and more broadly worldwide, with an emerging wind energy portfolio.

\if1\blind
{
\begin{center}
	\bf \Large Acknowledgments
\end{center}
This publication is based upon work supported by the King Abdullah University of Science and Technology (KAUST) Office of Sponsored Research (OSR) under Award No: OSR-2018-CRG7-3742.
}
\fi
\bibliographystyle{chicago}
\setlength{\bibsep}{5pt}
\bibliography{ref}

\clearpage

\setcounter{equation}{0}
\renewcommand{\theequation}{S\arabic{equation}}

\setcounter{figure}{0}
\setcounter{table}{0}
\setcounter{section}{0}
\setcounter{page}{1}
\renewcommand{\thesection}{S\arabic{section}}
\renewcommand{\thefigure}{S\arabic{figure}}
\renewcommand{\thetable}{S\arabic{table}}

\begin{center}
\bf \Huge  Supplementary Material
\end{center}

\section{Forecasts for Multiple Time-steps }
\label{sec:more-forecast}
We write $\bh_t(\bx^\ast_t)$ more explicitly as a function of $\bx^\ast_t$ in the forecasted $\hat\by^\ast_t$,
\[
\hat\by_t^\ast = 
\hat\bB\trans
\left[ 
\begin{array}{c}
\bh_t (\bx^\ast_t)\\
{\bh_t (\bx^\ast_t)\odot\bh_t (\bx^\ast_t)}
\end{array}
\right].
\]
At time $t$, the forecasted $\hat\by^\ast_{t+1}$ uses $\bx^\ast_{t+1}=\big(1,{\by^\ast_{t}}\trans,\ldots,{\by^\ast_{t-m+1}}\trans\big)\trans$;
the forecasted $\hat\by^\ast_{t+2}$ uses
\[
\bx^\ast_{t+2} = 
\left\{
\begin{array}{ll}
\big(1,\hat{\by}^\ast_{t+1}{}\trans,{\by^\ast_t}\trans\ldots,{\by^\ast_{t-m+2}}\trans\big)\trans &,~m > 1 \\
\big(1,\hat{\by}^\ast_{t+1}{}\trans\big)\trans &,~m = 1
\end{array}
\right.;
\]
and the forecasted $\hat\by^\ast_{t+3}$ uses
\[
\bx^\ast_{t+3} = 
\left\{
\begin{array}{ll}
\big(1,\hat{\by}^\ast_{t+2}{}\trans,\hat{\by}^\ast_{t+1}{}\trans,{\by^\ast_t}\trans\ldots,{\by^\ast_{t-m+3}}\trans\big)\trans &,~m > 2 \\
\big(1,\hat{\by}^\ast_{t+2}{}\trans,\hat{\by}^\ast_{t+1}{}\trans\big)\trans &,~m 
= 2 \\
\big(1,\hat{\by}^\ast_{t+2}{}\trans\big)\trans &,~m = 1
\end{array}
\right..
\]

\section{Discarded Forecasts at the Beginning and End of the Forecasting Period}
\label{subsec:discard}
Let the training period be $\{0,\ldots, T\}$ and the forecasting period be $\{T+1,\ldots,T_{\textrm{max}}\}$.
We write $\hat\by^\ast_{t\mid 1}$ as the one-hour-ahead forecast at time $t-1$, $\hat\by^\ast_{t\mid 2}$ as the two-hour-ahead forecast at time $t-2$, and $\hat\by^\ast_{t\mid 3}$ as the three-hour-ahead forecast at time $t-3$, respectively. The used time points for the forecasts are as follows:
\begin{itemize}
	\item $\hat\by^\ast_{t\mid1}: ~t\in\{T+1,\ldots,T_\textrm{max}\}$,
	\item $\hat\by^\ast_{t\mid2}: ~t\in\{T+2,\ldots,T_\textrm{max}\}$,
	\item $\hat\by^\ast_{t\mid3}: ~t\in\{T+3,\ldots,T_\textrm{max}\}$.
\end{itemize}
For example, we do not consider $\hat\by^\ast_{T+1\mid2}$ because the needed forecast, $\hat\by^\ast_{T\mid1}$, is not available.
At time $T_\textrm{max}-1$, we only forecast for one time-step, and the forecasts $\hat\by^\ast_{T_\textrm{max}+1\mid2}$ and $\hat\by^\ast_{T_\textrm{max}+2\mid3}$ are not required because we do not have data $\by^\ast_{T_\textrm{max}+1}$ and $\by^\ast_{T_\textrm{max}+2}$ for assessment.

\section{Simulation Study on a Modified Lorenz 96 Model}
\label{subsec:Lorenz}

We modify the Lorenz 96 model by adding a factor $\eta$  to control the nonlinearity in the dynamics as follows:
\begin{equation}\label{eq:modlor}
\frac{dy_i(t)}{dt} = \eta \big(y_{i+1}(t)-y_{i-2}(t)\big)y_{i-1}(t) - y_i(t) + F,
~i = 1, 2, \ldots, N,
\end{equation}
where $y_{-1}(t) \coloneqq y_{N-1}(t), y_0(t) \coloneqq y_N(t), y_{N+1}(t) \coloneqq y_1(t)$, $N$ is the number of sites, and $F$ is the external forcing.

We choose $N = 5$ and $F = 8$ and simulate data on $t\in [-199.9,100]$ by numerical integration with $\Delta t = 0.1$. The initial values are drawn from an independent standard normal distribution (i.e., $y_i(-199.9)
\overset{\text{i.i.d.}}{\sim}
N(0,1), i = 1,\ldots,N=5$). 
The realizations at the first 2,000 time points are treated as burn-in and discarded.
To account for the measurement error, we add white noise to the generated realizations (i.e., $\tilde y_i(t) = y_i(t) + \epsilon_i(t)$ where $ \epsilon_i(t) \overset{\text{i.i.d.}}{\sim}
N(0,1),~i = 1,\ldots,N=5,~t = 0.1, 0.2,\ldots,100$).
We proceed to model the simulated data $\tilde y_i(t),~i = 1,\ldots,N=5,~t=0.1,0.2,\ldots,100$.

We choose seven values, $0.2,0.4,\ldots,1.4$, for $\eta$ to control different levels of nonlinearity levels ($\eta$ is 1 in the original Lorenz 96 model).
Nonlinearity is more significant when $\eta$ is greater.
For each $\eta$, we generate 50 independent sets of realizations. We apply the ESN, Vector Autoregression (VAR), ARIMA, and persistence methods to the simulated data. 
Since we have a small number of sites ($N=5$), it is feasible to apply the VAR for the 5-variate time series. The VAR model is described as follows:
\[
\left[\begin{array}{c}
y_1(t)\\
\vdots\\
y_N(t)\\
\end{array}\right]
=
\left[\begin{array}{c}
c_1\\
\vdots\\
c_N\\
\end{array}\right]
+
\mathbf{A}_1
\left[\begin{array}{c}
y_1(t-1)\\
\vdots\\
y_N(t-1)\\
\end{array}\right]
+\cdots+
\mathbf{A}_p
\left[\begin{array}{c}
y_1(t-p)\\
\vdots\\
y_N(t-p)\\
\end{array}\right]
+
\left[\begin{array}{c}
\varepsilon_1(t)\\
\vdots\\
\varepsilon_N(t)\\
\end{array}\right]
,\]
where $p$ is the order of the vector autoregression term,
$\varepsilon_1(t),\ldots, \varepsilon_N(t)$ are error terms,
and $c_1,\ldots, c_N$ and the matrices $\mathbf{A}_1,\ldots, \mathbf{A}_p$ are unknown parameters.
It is noteworthy that the ESN and VAR model the realizations at the five sites altogether and make a 5-variate forecast at each time point, whereas the ARIMA models the time series at each site independently and performs univariate forecasting. 
All the parameters in these models are selected via the cross-validation by the MSE for the one-time-step forecast. 
In the cross-validation, the realizations at $t\in\{0.1,\ldots,50\}$ are used for the training, while those at $t\in\{50.1,\ldots,75\}$ are used for the validation.

\begin{figure}[ht!]
	\centering
	\includegraphics[width = \textwidth]{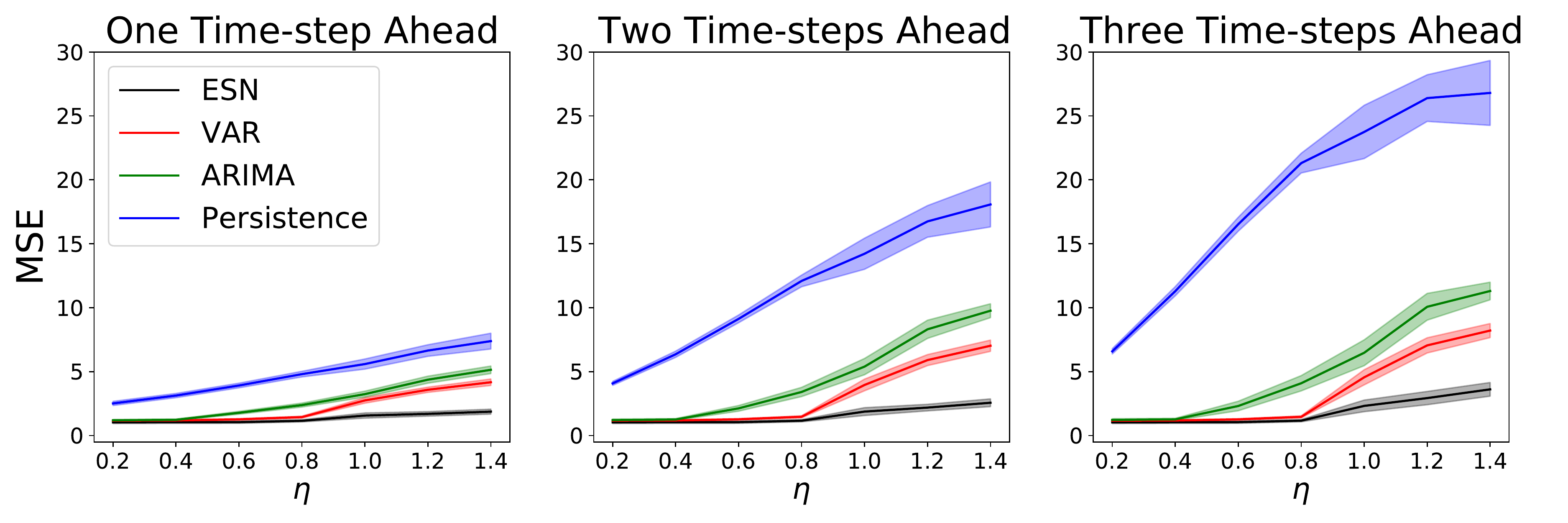}
	\caption{MSE by the ESN, VAR, ARIMA, and persistence models for the forecasts up to three time-steps ahead for $t\in\{50.1,\ldots,75\}$ for the simulated data from the modified Lorenz 96 model in Equation~\eqref{eq:modlor}. The solid lines show the mean MSE in the 50 experiments, whereas the bands indicate the mean $\pm$ the standard deviation.}
	\label{fig:lorenz}
\end{figure}

Once the optimal parameter values are obtained, we compute forecasts up to three time-steps ahead for $t\in\{75.1,\ldots,100\}$ by each method.
Figure~\ref{fig:lorenz} illustrates the MSE results.
Short-term forecasting is more difficult when the nonlinearity of the dynamic becomes more significant as $\eta$ increases. 
It is no surprise that forecasts for more time-steps ahead are less accurate.
However, in all cases, the ESN leads to the most accurate forecasts. 
When $\eta$ is less, the outperformance of the ESN is not obvious. When $\eta$ departs from zero, the ESN is generally more capable of capturing the nonlinear dynamics compared to other linear methods.

\clearpage
\section{Supplementary Figures}

\begin{figure}[ht!]
	\centering
	\includegraphics[width=\linewidth]{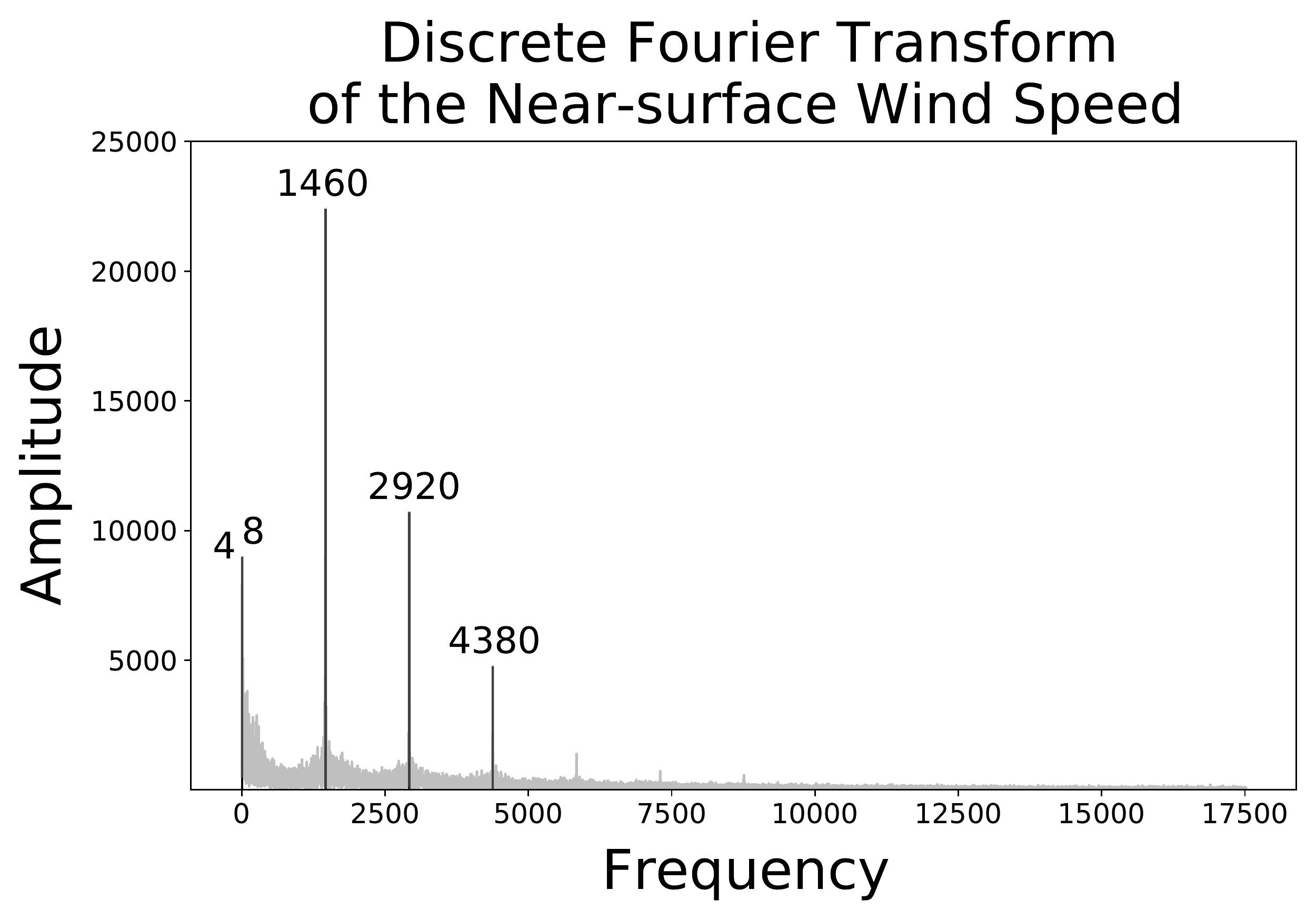}
	\caption{Discrete Fourier transform of the near-surface wind speed. Highlighted are the significant frequencies corresponding to periods of one year, half a year, one day, twelve hours, and eight hours.}
	\label{fig:DFT}
\end{figure}

\begin{figure}[ht!]
	\centering
	\includegraphics[width=\linewidth]{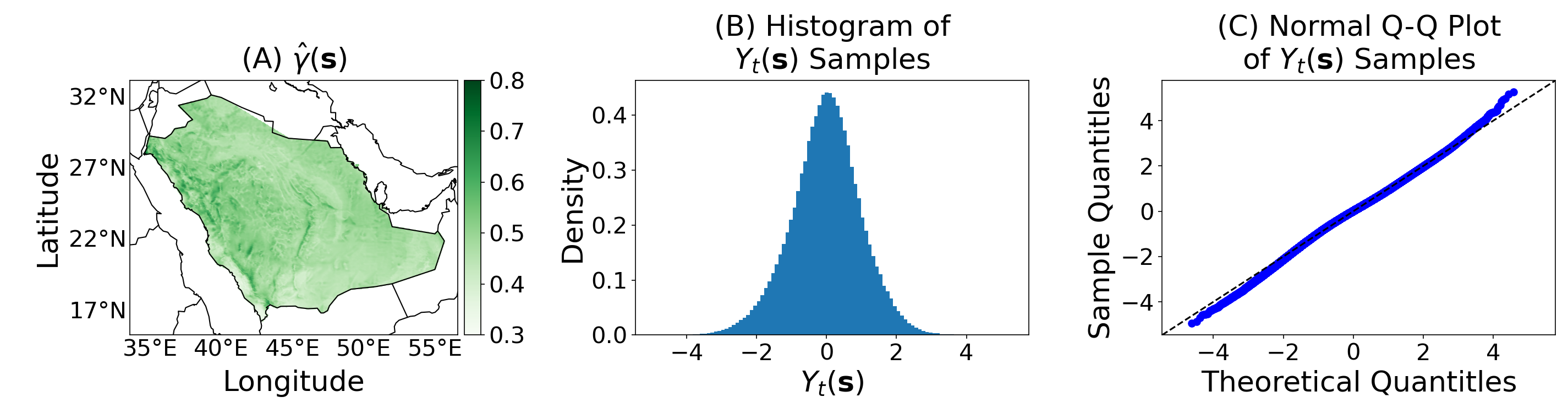}
	\caption{(A): Scaling parameter estimate $\hat\gamma(\bs)$ at each location. (B): Histogram of $Y_t(\bs)$ samples over a regular space-time grid. (C): The samples' Q-Q plot for a normal distribution.}
	\label{fig:qqplot}
\end{figure}

\begin{figure}[ht!]
	\centering
	\includegraphics[width = \textwidth]{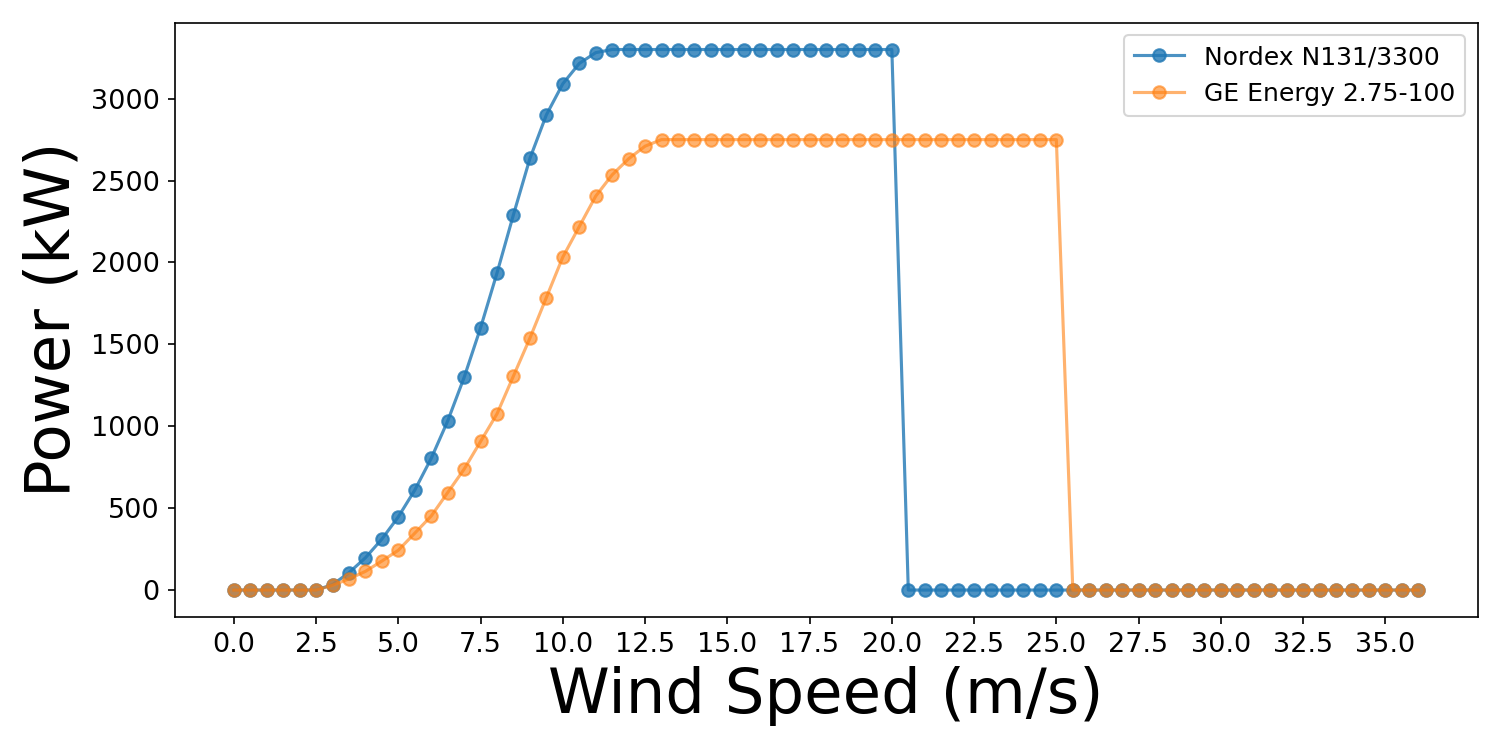}
	\caption{Power curve provided by manufacturer for transforming wind speed to power.}
	\label{fig:power_curve}
\end{figure}

\clearpage
\section{Supplementary Tables}

\begin{table}[ht!]
	\centering
	\caption{\label{tab:search_grid}Search grid for the parameters in the cross-validation.}
	\vspace{3mm}
	\begin{tabular}{c|c}
		\hline
		Parameters & Grid \\ \hline \hline
		$n_h$ & $\{1000,1500,\ldots,5000\}$ \\ \hline
		$m$ & $\{1,2,\ldots,10\}$ \\ \hline
		$\phi$ & $\{0.1,0.2,\ldots,1\}$ \\ \hline
		$\delta$ & $\{0.05,0.1,0.15,\ldots,2\}$ \\ \hline
		$\lambda$ & $\{0.01,0.05,0.1,0.15,0.2,0.25,0.3\}$ \\ \hline
		$a_w$ & $\{0.005,0.01,0.05,0.1,0.15\}$ \\ \hline
		$a_u$ & $\{0.005,0.01,0.05,0.1,0.15\}$ \\ \hline
		$\pi_w$ & $\{0.005,0.01,0.05,0.1,0.15\}$ \\ \hline
		$\pi_u$ & $\{0.005,0.01,0.05,0.1,0.15\}$ \\ \hline
	\end{tabular}
	
\end{table}

\begin{table}[ht!]
	\centering
	\caption{\label{tab:UQ-all} The mean proportion of the true wind speed residuals $Y_t(\bs)$ at all the $n$ locations in 2016 falling into the associated prediction intervals (standard deviation across the locations is shown in parentheses).}
	\vspace{3mm}
	\begin{tabular}{|c|c|c|c|}
		\hline
		Prediction &  \multicolumn{3}{c|}{Prediction Interval Coverage} \\
		\cline{2-4}
		Interval & One hour ahead & Two hours ahead & Three hours ahead \\
		\hline
		$95\%$  &  $94.6\% (0.7\%)$ &  $94.5\% (0.7\%)$ &  $94.5\% (0.7\%)$\\
        $80\%$  &  $80.5\% (1.1\%)$ &  $80.6\% (1.1\%)$ &  $80.6\% (1.2\%)$\\
        $60\%$  &  $60.1\% (1.5\%)$ &  $60.1\% (1.4\%)$ &  $60.0\% (1.5\%)$\\
		\hline
	\end{tabular}
	
\end{table}
\end{document}